\documentclass[conference]{IEEEtran}
\IEEEoverridecommandlockouts

\usepackage{amsmath,amssymb,amsfonts}

\usepackage{graphicx}
\usepackage{array}
\usepackage{tabularx}

\usepackage{textcomp}
\usepackage{xcolor}
\usepackage{dsfont}

\usepackage{algorithm}
\usepackage{algpseudocode}
\usepackage{cite}
\usepackage{hyperref}
\usepackage{etoolbox}
\usepackage{comment}

\def\BibTeX{{\rm B\kern-.05em{\sc i\kern-.025em b}\kern-.08em
    T\kern-.1667em\lower.7ex\hbox{E\kern-.125emX}}}

\begin{document}
\title{ Network Adaptation in IRS-Aided Hybrid RF/VLC Systems Using Cooperative Multi-Agent DRL}
\author{\IEEEauthorblockN{Ahrar N. Hamad, Ahmad Adnan Qidan, Taisir E.H. El-Gorashi and Jaafar M. H. Elmirghani, }
\IEEEauthorblockA{\textit{ Department of Engineering, King’s College London, London, United Kingdom,} \\
\textit
\{{ahrar.hamad, ahmad.qidan, taisir.elgorashi, jaafar.elmirghani}\}@kcl.ac.uk}
}
\maketitle
\begin{abstract} Hybrid radio frequency (RF) and visible light communication (VLC) networks have emerged as a promising solution for high-capacity indoor wireless connectivity in sixth-generation (6G) systems. However, the limited optical coverage and vulnerability of VLC links to line-of-sight (LoS) blockage under user mobility remain fundamental challenges. In this work, a mirror-based intelligent reflecting surface (IRS) is deployed to assist a dynamic indoor hybrid RF/VLC network, where each mobile user is exclusively assigned to either the IRS-enhanced VLC subnetwork or the RF subnetwork through a binary selection decision. A joint optimization problem is then formulated to maximize proportional fairness by jointly optimizing the RF/VLC technology selection, power allocation, and IRS mirror roll and yaw orientation angles. To enable real-time adaptability, the problem is reformulated as a Markov decision process (MDP) and solved using a cooperative multi-agent deep reinforcement learning (DRL) algorithm based on centralized training with decentralized execution. Simulation results demonstrate the superior performance of the optimized hybrid network compared with optimized standalone VLC and RF networks. The results further validate the practicality and effectiveness of the proposed DRL framework compared to widely adopted DRL algorithms as well as conventional model-based optimization approaches.
\end{abstract}
\section{Introduction}\label{Sec:Intro}
Visible light communication (VLC) is an attractive option for simultaneously delivering high-speed data transmission and illumination through optical access points (APs), while reusing the existing lighting infrastructure to align with the green communication objectives of sixth-generation (6G) networks. It offers access to the unregulated optical spectrum, high spectral efficiency, and complete immunity to radio-frequency (RF) interference. Therefore, VLC can be integrated with conventional RF networks to form hybrid RF/VLC systems, offering higher network capacity, enhanced reliability, and improved energy efficiency \cite{9064520}. However, VLC systems still face fundamental coverage limitations, as the coverage footprint of optical APs is spatially constrained, while line-of-sight (LoS) links are highly susceptible to blockage by physical objects and user mobility \cite{11428181,11673253}. Intelligent reflecting surface (IRS) has emerged as an effective solution for enhancing connectivity in indoor VLC systems without consuming significant power \cite{10972182}. Specifically, a steerable IRS mirror array can overcome LoS blockage by dynamically redirecting non-LoS (NLoS) optical signals toward mobile users, effectively extending the optical AP coverage footprint and improving the achievable data rate \cite{11370966}.

Hybrid RF/VLC networks introduce complex optimization challenges involving technology selection and resource allocation. In \cite{10034535}, alternating optimization was employed to jointly optimize user association and power allocation in a hybrid RF/VLC system using successive convex approximations (SCA). Similarly, in \cite{9411734}, a joint AP assignment and power allocation problem was formulated as a combinatorial optimization problem to maximize the sum rate and solved using a matching theory-based distributed algorithm combined with a heuristic power allocation scheme. However, these approaches treat the VLC channel as fixed and uncontrollable, limiting their ability to adapt to dynamic propagation conditions. Building on the controllable NLoS capability enabled by IRSs, the work in \cite{aboagye2021intelligent} employed a sine-cosine algorithm to determine the optimal IRS mirror configuration for establishing reliable NLoS links. Nevertheless, such model-based optimization approaches typically rely on accurate channel knowledge and require repeated optimization whenever channel conditions change, making them less suitable for real-time adaptation in dynamic environments with user mobility.

Deep reinforcement learning (DRL) has recently emerged as an effective framework for solving optimization problems including integer variables, coupled continuous actions, and non-convex objectives \cite{sutton2018reinforcement}. Unlike model-based methods, a DRL agent learns an adaptive policy through interaction with the environment, enabling real-time response to mobility and blockage. In \cite{RF/VLC}, a deep Q-network (DQN) algorithm jointly optimized user association, bandwidth, and power in a hybrid RF/VLC network. However, its discrete action space cannot capture continuous control variables. In \cite{10844503}, such challenges were addressed using on-policy algorithms for resource allocation and load balancing in hybrid RF/VLC systems. In \cite{saifaldeen2024drl}, a deep deterministic policy gradient (DDPG) algorithm is used to optimize the transmit power among the users in the hybrid RF/VLC system. Despite these advances, existing works do not jointly consider IRS reconfiguration, technology selection, and power allocation, particularly in dynamic IRS-aided hybrid RF/VLC scenarios involving user mobility. 

\begin{figure}[t]
    \centering
    \includegraphics[width=0.47\textwidth]{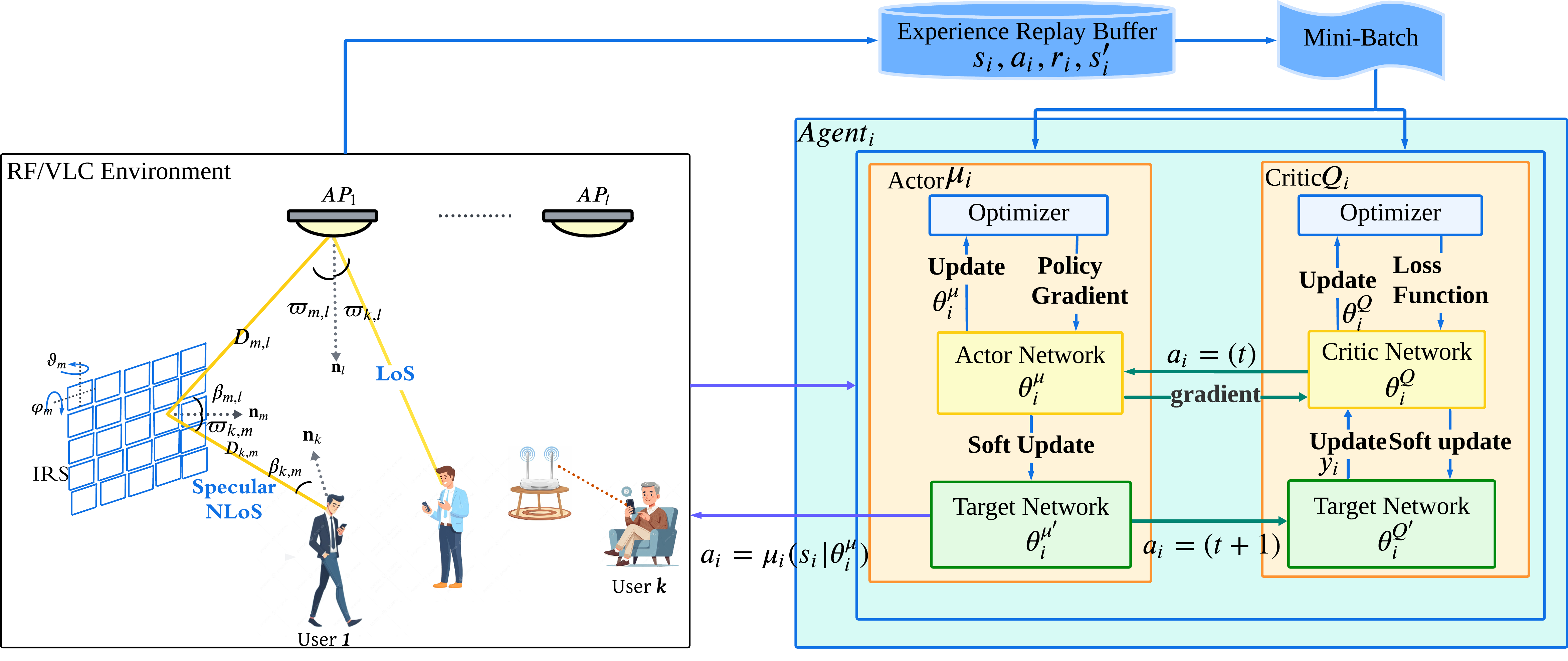}
    \caption{IRS-assisted RF/VLC system model with DRL framework.}
    \vspace{-4mm}
    \label{fig:System_model}
\end{figure}
Motivated by these gaps, this paper proposes a novel framework for jointly optimizing RF/VLC link selection, power allocation, and IRS mirror orientation in a dynamic multi-user indoor environment. The resulting optimization problem is formulated as a Markov decision process (MDP) and addressed using a multi-agent DRL framework based on centralized training with decentralized execution. This enables the system to continuously adapt to user mobility and blockage while maximizing proportional fairness among users. The proposed framework is designed to operate in real-time and effectively handle the hybrid discrete-continuous action space and non-convex nature of the IRS-assisted RF/VLC optimization problem. Simulation results demonstrate that our proposed approach consistently outperforms state-of-the-art benchmark schemes.
\vspace{-4mm}
\section{System Model}\label{Sec:system}

We consider a downlink indoor IRS-assisted hybrid RF/VLC system, as shown in Fig. \ref{fig:System_model}. A number of light-emitting diode (LED) APs $l = \{1, \ldots,N_l\}$ are placed on the ceiling for simultaneous communication and illumination, while the RF APs $f = \{1, \ldots,N_f\}$ are deployed on the walls to provide complementary wireless coverage. On the communication floor, a set of mobile users $k = \{1, \ldots, K\}$, move at different velocities following random waypoint (RWP) mobility patterns. Each user $k$ is equipped with a multi-branch angle diversity receiver for VLC reception and an RF antenna, enabling dynamic technology selection. Each user connects to either the VLC subnetwork or the RF subnetwork, but not both simultaneously. This selection is determined by a binary variable $\alpha_k \in \{0,1\}$, where $\alpha_k = 1$ indicates association with the VLC subnetwork and $\alpha_k = 0$ corresponds to the RF subnetwork. Furthermore, an IRS mirror array is mounted on one of the walls and consists of $\mathcal{M} = \{1, \ldots, M\}$ mirrors. The area of each mirror is denoted by $dA_m = h_m \times w_m$, where $h_m$ and $w_m$ represent the height and width of mirror $m$, respectively. A spacing $q_m$ is considered between adjacent mirrors to avoid interference during rotation. In addition, the orientation of each rotational mirror can be controlled by two angles: the roll angle $\varphi_m$ and the yaw angle $\vartheta_m$, using micro-electro-mechanical systems (MEMS) technology \cite{11370966}. This rotation allows each mirror to steer the reflected VLC beam toward intended users and effectively extends the coverage of the VLC subnetwork. A central controller unit coordinates both the VLC and RF subnetworks by collecting channel state information from all APs and users and determines the optimal technology selection for each user.

\vspace{-2mm}
\subsection{VLC Channel Model}
The total VLC channel gain $h_{k,l}^{(v)}$ from LED AP $l$ to user $k$ consists of two components: the direct LoS path  $h_{k,l}^{\text{LoS}}$ and the IRS-reflected NLoS path $h_{k,l,m}^{\text{IRS}}$.  In practical indoor environments, the LoS optical link between LED AP $l$ and user $k$ is frequently obstructed by physical objects such as furniture, partitions, and the users themselves. To capture this effect, the LoS blockage is modeled as an independent Bernoulli random process. At each transmission time slot, the LoS link between AP $l$ and user $k$ is available with probability $1-p_b$ and blocked with probability $p_b$,
where $p_b$ denotes the blockage probability. Accordingly, the total VLC channel gain is expressed as

{\small
\begin{equation}
h_{k}^{(v)} = \sum_{l=1}^{N_l}\!\left(b_{k,l}\,
h_{k,l}^{\mathrm{LoS}} + \sum_{m=1}^{M}
h_{k,m,l}^{\mathrm{IRS}}(\varphi_m, \vartheta_m)\right),
\label{eq:h_vlc_blockage}
\end{equation}}

\noindent where $b_{k,l} \in \{0,1\}$ is the blockage indicator with $\Pr(b_{k,l}=0) = p_b$. Note that the IRS-reflected NLoS component remains available even when the direct LoS path is blocked, provided the mirror orientations are appropriately configured. This highlights the critical role of IRS in maintaining VLC connectivity under blockage conditions. In this work, the direct LoS component $h_{k,l}^{\text{LoS}}$ follows the Lambertian radiation model and is given by
\vspace{-3mm}

{\small 
\begin{equation}
\begin{aligned}
  h_{k,l}^{LoS}
  &= \frac{(m_L + 1)A_{PD}T_f\, g(\psi_{k,l})}
{2\pi d_{k,l}^2}\cos^{m_L}(\varpi_{k,l})\cos(\psi_{k,l}),
\\ & 0 \leq \psi_{k,l} \leq \psi_c 
\end{aligned}
\vspace{-4pt}
\end{equation}}

\noindent where $m_L = -\frac{\ln 2}{\ln(\cos \Phi_{1/2})}$ is the Lambertian emission order, with LED half-power radiation angle $\Phi_{1/2}$. $A_{\text{PD}}$ is the physical area of the photodetector,$T_f$ is the optical filter gain, $g(\psi_{k,l})$ is the optical concentrator gain, $d_{k,l}$ is the Euclidean distance between AP $l$ and user $k$, and $\varpi_{k,l}$ and $\psi_{k,l}$ are the irradiance and incidence angles, respectively. $\psi_c$ is the photodetector semi-angle field of view.

The channel gain $h_{k,m,l}^{\text{IRS}}$ of the indirect signal is reflected by the IRS mirror $m$ with complete specular reflection and orientation angles $(\varphi_m,\vartheta_m)$, is given by \cite{10972182,11370966}

{\small
\begin{equation}
\label{IRSCH}
\begin{aligned}
h_{k,m,l}^{\text{IRS}}
={}&
\frac{(m_L+1)\,\rho_m A_{PD}\, dA_m
\cos^{m_L}(\varpi_{m,l})\cos(\beta_{m,l})}
{2\pi^2 (d_{m,l}+d_{k,m})^2}
\\
&\times
\cos(\varpi_{k,m})\cos(\beta_{k,m}),
\qquad
0\le\beta_{k,m}\le\psi_c
\end{aligned}
\end{equation}}

\noindent where $\rho_m$ is the mirror reflection coefficient, $d_{m,l}$ and $d_{k,m}$ are the distances from LED AP $l$ to mirror $m$ and from mirror $m$ to user $k$, respectively. \( \varpi_{m,l} \) is the radiation angle from AP \( l \) to mirror \( m \), $\beta_{m,l}$ is the incidence angle from AP $l$ to mirror $m$. Moreover, $\varpi_{k,m}$ is the irradiance angle from mirror $m$ towards user $k$, and \( \beta_{k,m} \) is the incidence angle of the signal reflected from mirror \( m \) to user \( k \), which are explicit functions of the roll angle $\varphi_m$ and yaw angle $\vartheta_m$ through the mirror surface normal vector. The total received signal at user $k$ assigned to the VLC subnetwork (i.e., $\alpha_k = 1$) is given by
\vspace{-3mm}

{\small
\begin{equation}
\label{eq:vlc_signal}
y_k^{(v)} = \underbrace{\eta \rho^{(v)} P_k^{(v)}
h_k^{(v)} s_k}_{\text{desired}}
+ \underbrace{\eta \!\!\sum_{\substack{j=1\\j\neq k}}^{K}
\!\!\alpha_j \rho^{(v)} P_j^{(v)} h_k^{(v)} s_j}
_{\text{interference}}
+ n_k^{(v)},
\end{equation}}
\vspace{-4mm}

\noindent where $\eta$ is the photodetector responsivity, $\rho^{(v)}$ is the total optical transmit power per LED AP, $P_j^{(v)} \in [0,1]$ is the normalized power fraction allocated to user $j$, $s_j$ is the data symbol intended for user $j$ 
and $n_k^{(v)} \sim \mathcal{N}(0,\sigma_o^2)$ is the additive white Gaussian noise (AWGN) at the photodetector. The interference term includes only VLC-assigned users, as enforced by the selection variable $\alpha_j$. For user $k$ assigned to VLC, 
the signal-to-interference-plus-noise ratio (SINR) $\gamma_k^{(v)}$ is given by 
\vspace{-6mm}

{\small
\begin{equation}
\label{eq:vlc_sinr}
\gamma_k^{(v)} =
\frac{\left(\eta \rho^{(v)} P_k^{(v)} h_k^{(v)}\right)^2}
{\displaystyle\sum_{\substack{j=1\\j\neq k}}^{K}
\alpha_j\left(\eta \rho^{(v)} P_j^{(v)}
h_k^{(v)}\right)^2 + \sigma_o^2},
\end{equation}}

\noindent where the squared terms arise from the intensity modulation and direct detection (IM/DD) \cite{11673253}. The data rate of user $k$ in the VLC subnetwork is given by 

{\small
\begin{equation}
R_k^{(v)} = {B_w^{(v)}} \log_2 \left(1 + \frac{e}{2\pi} \gamma^{(v)}_k \right),
\end{equation}}
\vspace{-5mm}

\noindent where $B_w^{(v)}$ represents the VLC system modulation bandwidth and ${e}$ accounts for the Euler number. 
\vspace{-1mm}

\subsection { RF Channel Model}
The channel from RF AP $f$ to user $k$ follows a Rician fading model as follows
{\small
\begin{equation}
h_{k,f}^{(r)} =
\sqrt{\beta_{k,f}^{(r)}}
\left(
\sqrt{\frac{\kappa}{\kappa + 1}}\, e^{j\theta_{k,f}}
+
\sqrt{\frac{1}{\kappa + 1}}\, g_{k,f}
\right),
\end{equation}}
\vspace{-3mm}

\noindent where $\beta_{k,f}^{(r)}$ represents the large-scale fading coefficient between RF AP $f$ and user $k$ at distance $d_{k,f}$, given by
$\beta_{k,f}^{(r)} = \beta_0 \left(\frac{d_0}{d_{k,f}}\right)^{\mu},$ where $\beta_0$ is the path loss at reference distance $d_0$, and $\mu$ is the path loss exponent. The parameter $\kappa$ is the Rician $\kappa$-factor representing the ratio of LoS to NLoS power, $\theta_{k,f}$ is the phase angle of the LoS component, and  $g_{k,f}$ is the non-LoS component. The received signal at user $k$ assigned to the RF subnetwork (i.e., $\alpha_k = 0$) is given by \cite{9064520}
\vspace{-7pt}

\begin{equation}
\resizebox{\linewidth}{!}{$
\begin{aligned}
y_k^{(r)}
={}&
\underbrace{\sum_{f=1}^{N_f} h_{k,f}^{(r)}
\sqrt{\rho^{(r)}P_k^{(r)}}\,s_k}_{\text{desired}}
+
\underbrace{\sum_{\substack{j=1\\j\neq k}}^{K}
\mathcal{W}\sum_{f=1}^{N_f} h_{k,f}^{(r)}
\sqrt{\rho^{(r)}P_j^{(r)}}\,s_j}_{\text{interference}}
+n_k^{(r)},
\end{aligned}
$}
\end{equation}
\vspace{-2mm}

\noindent where $\mathcal{W}=(1-\alpha_j)$, $\rho^{(r)}$ is the total available transmit power at each RF AP $f$, $P_k^{(r)}$ is the normalized power fraction allocated to user $k$ in the RF subnetwork, $s_k$ is the data symbol intended for user $k$ and $n_k^{(r)} \sim \mathcal{CN}(0,\sigma_r^2)$ is the AWGN at the RF receiver. 
Moreover, for user $k$ assigned to RF (i.e., $\alpha_k = 0$), the received SINR $\gamma_k^{(r)}$ is given by
\vspace{-3mm}

{\small
\begin{equation}
\label{eq:sinr}
\gamma_k^{(r)}=
\frac{
\rho^{(r)}P_k^{(r)}
\left|h_k^{(r)}\right|^2
}{
\displaystyle
\sum_{\substack{j=1\\j\neq k}}^{K}
(1-\alpha_j)
\rho^{(r)}P_j^{(r)}
\left|h_k^{(r)}\right|^2
+\sigma_r^2
}.
\end{equation}}

\noindent The achievable rate of user $k$ in the RF subnetwork is given by
\vspace{-2mm}
{\small
\begin{equation}
R_k^{(r)} = B^{(r)} \log_2 \left(1 + \gamma_k^{(r)} \right), 
\end{equation}}
\vspace{-3mm}

\noindent where $B^{(r)}$ denotes the RF bandwidth. Finally, since each user is associated with either the VLC subnetwork or the RF subnetwork, the available power and bandwidth resources of each subnetwork are allocated exclusively to its associated users. Therefore, the achievable rate of user $k$ is given by
\vspace{-2mm}
\begin{equation}
\label{TDR}
R_k = \alpha_k R_k^{(v)} + (1 - \alpha_k) R_k^{(r)}.
\end{equation}

\section{Problem Formulation}\label{Sec:ProbForm}
\vspace{-1mm}
We now formulate a joint optimization problem to determine the optimal RF/VLC technology selection, power allocation across both subnetworks, and IRS mirror orientations, 
with the objective of maximizing the proportional fairness utility, which balances aggregate sum rate and inter-user fairness. The optimization problem is formulated as follows
\vspace{-2mm}

{\small
\begin{subequations}\label{P1}
\begin{align}
\textbf{P1:} \quad
\max_{\boldsymbol{\alpha},\, \mathbf{P}^{(v)},\, \mathbf{P}^{(r)},\,
\boldsymbol{\varphi},\, \boldsymbol{\vartheta}}
\quad & \sum_{k=1}^{K} \log(R_k)
\tag{12} \label{P1_obj} \\
\text{s.t.} \quad
&  R_k \geq R_{\min,k} ,
  \quad \forall k \in \mathcal{K} \label{C1} \\
& \sum_{k=1}^{K} \alpha_k P_k^{(v)} \leq \rho^{(v)} ,\label{C2} \\
& \sum_{k=1}^{K} (1-\alpha_k) P_k^{(r)} \leq \rho^{(r)}, \label{C3} \\
&  0 \leq P_k^{(\xi)} \leq 1,
  \quad \forall \xi \in \{v,r\},\, \forall k \in \mathcal{K} \label{C4} \\
& \alpha_k \in \{0,1\}, 
  \quad \forall k \in \mathcal{K} \label{C5} \\
& \frac{-\pi}{2} \leq \varphi_m \leq \frac{\pi}{2}, 
  \quad \forall m \in \mathcal{M} \label{C6} \\
&  \frac{-\pi}{2} \leq \vartheta_m \leq \frac{\pi}{2}, 
  \quad \forall m \in \mathcal{M} \label{C7}
\end{align}
\end{subequations}}

\noindent where constraint \eqref{C1} ensures the minimum quality-of-service (QoS) requirement. Constraint \eqref{C2} limits the total power allocated to VLC-associated users to the available power budget $\rho^{(v)}$ at the LED AP, while \eqref{C3} similarly limits the total transmit power allocated to RF-associated users to the available RF power budget $\rho^{(r)}$. Constraint \eqref{C4} bounds the normalized power-allocation coefficients for both subnetworks, and \eqref{C5} enforces the binary nature of the technology-selection variable $\alpha_k$. Constraints \eqref{C6} and \eqref{C7} set the physical rotation limits of the  IRS mirrors. The resulting problem $\mathbf{P1}$ is a non-convex mixed-integer nonlinear program. The non-convexity arises from the binary technology-selection variables $\alpha_k$, their bilinear coupling with the power-allocation variables in \eqref{C2} and \eqref{C3}, and the nonlinear dependence of the VLC achievable rate $R_k^{(v)}$ on the IRS mirror orientations. Conventional methods such as exhaustive search and successive convex approximation (SCA) are unsuitable for real-time environment \cite{11673253,11164782}, as they must be re-solved from scratch for every channel realization, which is computationally prohibitive under the user mobility and blockage conditions considered in this work, where the channel varies over time. To enable efficient real-time decision-making, $\mathbf{P1}$ is therefore reformulated as an MDP and solved using our proposed intelligent DRL framework described in the following section.

\section{Proposed Intelligent DRL Framework }\label{Sec:RL}

The optimization problem in \eqref{P1_obj} is reformulated as an MDP characterized by the tuple $(\mathcal{S}, \mathcal{A}, \mathcal{R}, \mathcal{P}, \gamma)$, where $\mathcal{S}$ is the global state space, $\mathcal{A}$ is the joint action space of both agents, $\mathcal{R}$ is the shared cooperative reward encoding the objective of \textbf{P1}, $\mathcal{P}$ is the state transition probability under the joint actions of the agents, and $\gamma \in (0,1)$ is the discount factor balancing immediate and long-term rewards. The hybrid discrete-continuous structure of \textbf{P1} cannot be efficiently accommodated by conventional DRL approaches. Value-based methods such as DQN \cite{RF/VLC} operate only over discrete action spaces and would require quantizing the continuous power fractions and mirror angles, causing a combinatorial explosion that is intractable for practical system sizes. Policy-gradient methods such as DDPG \cite{saifaldeen2024drl} support continuous actions, but a single agent must learn a policy spanning the binary selection variables $\boldsymbol{\alpha}$, the power fractions, and the mirror angles within one heterogeneous action space, which couples variables operating on different scales and physical timescales and degrades convergence stability. Motivated by these limitations, the proposed framework decomposes the joint action space across cooperative agents: the selection and power agent $AG_{SP}$ operates on the scheduling variables $\{\boldsymbol{\alpha}, \mathbf{P}^{(v)}, \mathbf{P}^{(r)}\}$ executed by the central controller, while the IRS orientation agent $AG_{IRS}$ operates on the mirror angles $\{\boldsymbol{\varphi}, \boldsymbol{\vartheta}\}$ executed by the IRS controller. The agents follow a natural two-timescale ordering in which $AG_{IRS}$ acts first to reshape the VLC channel, after which $AG_{SP}$ observes the updated channel and makes an informed selection decision. To preserve coordination, the agents are trained under centralized training with decentralized execution, where each critic accesses the joint state and both agents' actions during training while each actor relies only on its local observation at execution.

The state space $\mathcal{S}$ consists of the VLC channel gain vector $\mathbf{h}_t^{(v)} \in \mathbb{R}^{K}$, the RF channel gain vector $\mathbf{h}_t^{(r)} \in \mathbb{C}^{K}$, the vector of achievable user rates $\mathbf{R}_{t-1} \in \mathbb{R}^{K}$, the user association vector $\boldsymbol{\alpha}_{t-1} \in \{0,1\}^{K}$, the spatial coordinates of all users $\mathbf{u}_t \in \mathbb{R}^{K \times 2}$, and the IRS roll and yaw angle vectors $\boldsymbol{\varphi}_{t-1} \in \mathbb{R}^{M}$, $\boldsymbol{\vartheta}_{t-1} \in \mathbb{R}^{M}$. The action space $\mathcal{A}$ contains all the actions, which are the outputs of the actor networks of both agents at each time slot $t$. Each agent $AG_i \in \{AG_{SP},\, AG_{IRS}\}$ selects an action $a_t \in \mathcal{A}$ based on its local observation $o_i$ and must satisfy the constraints in \eqref{P1}. The first agent $AG_{SP}$ selects action $a_{SP,t} \in \mathcal{A}$ to jointly determine the binary RF/VLC technology assignment $\alpha_{k,t}$ and the power allocation fractions $P_{k,t}^{(v)}$ and $P_{k,t}^{(r)}$ for each user $k \in \mathcal{K}$. The selected action must satisfy the selection constraint \eqref{C5} and the power constraints \eqref{C2}-\eqref{C4}, and is given by
\vspace{-2mm}

{\small
\begin{equation}
    a_{SP,t} = \Big[\, \alpha_{k,t},\, P_{k,t}^{(v)},\, 
    P_{k,t}^{(r)} \;\Big|\; k \in \mathcal{K} \,\Big].
    \label{eq:action_sp}
\end{equation}}

The second agent $AG_{IRS}$ selects action $a_{IRS,t} \in \mathcal{A}$ to determine the roll angle $\varphi_{m,t}$ and yaw angle $\vartheta_{m,t}$ of each IRS mirror $m \in \mathcal{M}$, allowing the IRS to steer the reflected VLC signals toward the intended users. This action must satisfy the mirror orientation constraints \eqref{C6} and \eqref{C7}, and is given by
\vspace{-2mm}

{\small
\begin{equation}
    a_{IRS,t} = \Big[\, \varphi_{m,t},\, \vartheta_{m,t} 
    \;\Big|\; m \in \mathcal{M} \,\Big].
    \label{eq:action_irs}
\end{equation}}

The reward function is designed to directly encode the proportional fairness objective of \textbf{P1} while penalizing violations of the QoS constraint~\eqref{C1} and the power budget constraints~\eqref{C2} -\eqref{C3}. Both agents receive the same shared cooperative reward $r_t$ at each time step $t$, determined by the current state $s_t$ and the joint action $a_t = \{a_{SP,t}, a_{IRS,t}\}$. The use of a shared reward encourages both agents to coordinate their actions toward the common goal of maximizing proportional fairness across all users and is given by
\vspace{-3mm}

{\small
\begin{equation}
\label{eq:reward}
\begin{aligned}
r_t ={}&
\sum_{k=1}^{K}\log(R_{k,t})
-\lambda_1\sum_{k=1}^{K}
\max\!\left(0,R_{\min,k}-R_{k,t}\right)
\\
&-\lambda_2\!\left[
\max\!\left(0,\mathcal{P}_t^{(v)}-\rho^{(v)}\right)
+\max\!\left(0,\mathcal{P}_t^{(r)}-\rho^{(r)}\right)
\right],
\end{aligned}
\end{equation}}

\noindent where $\mathcal{P}^{(v)}_t = \sum_{k=1}^{K}\alpha_{k,t} P_{k,t}^{(v)}$ and $\mathcal{P}^{(r)}_t = \sum_{k=1}^{K}(1-\alpha_{k,t})P_{k,t}^{(r)}$ are the total allocated VLC and RF power at time slot $t$, respectively. $\lambda_1 > 0$ penalizes violations of the minimum QoS constraint~\eqref{C1}, while $\lambda_2 > 0$ penalizes violations of the power budget constraints~\eqref{C2} and~\eqref{C3}. The $\max(0,\cdot)$ ensures that only active constraint violations contribute to the penalty, so that the reward is not unnecessarily reduced when all constraints are satisfied.


For better understanding, the proposed algorithm training procedure is provided in Algorithm~\ref{alg:maddpg}. At each time step $t$, $AG_{IRS}$ acts first to reconfigure the IRS mirror orientations, which immediately updates the VLC channel gains for all users. $AG_{SP}$ then observes the updated channel state and determines the technology assignment and power allocation accordingly. The joint action $a_t = \{a_{SP,t}, a_{IRS,t}\}$ is executed in the environment, the per-user rates $R_{k,t}$ are computed, and the shared cooperative reward $r_t$ is evaluated via~\eqref{eq:reward}, after which the next state $s_{t+1}$ is observed. The transition tuples $\{s_t, a_t, r_t, s_{t+1}\}$ are stored in the replay buffer $\mathcal{D}$ and randomly sampled in mini-batches of size $\mathcal{N}_b$ to update each agent's critic by minimizing the temporal difference loss, the actor using the deterministic policy gradient, and both target networks using the soft update rule. This proposed framework jointly and adaptively optimizes network performance while satisfying constraints \eqref{C1}-\eqref{C7} at each time slot as user locations and channel conditions change.

The computational complexity is analyzed for both training and execution. Let $L_\mu$, $n_\mu$ and $L_Q$, $n_Q$ denote the number of layers and widest-layer neurons of the actor and critic networks, respectively. Per training iteration, each agent requires $\mathcal{O}(L_\mu n_\mu^2)$ for the actor update, $\mathcal{O}(L_Q n_Q^2)$ for the critic, and $\mathcal{O}(n_\mu + n_Q)$ for the target networks, plus $\mathcal{O}(K N_l M)$ for the VLC channel update following IRS reconfiguration. Over $N_{ep}$ episodes of $T$ steps with mini-batch size $\mathcal{N}_b$, and accounting for all agents,  the total training complexity is $\mathcal{O}(N_{ep} T \mathcal{N}_b (L_\mu n_\mu^2 + L_Q n_Q^2))$, with space complexity dominated by the replay buffer $\mathcal{O}(|\mathcal{D}|(|\mathcal{S}| + |\mathcal{A}|))$. As training is performed offline, the per-slot execution cost reduces to a single forward pass through each actor, $\mathcal{O}(L_\mu n_\mu^2 + K N_l M)$, which is independent of the channel realization. In contrast, SCA requires $\mathcal{O}(I_{\mathrm{SCA}} (K+M)^3)$ operations per channel realization, since each iteration solves a convex subproblem jointly over the selection, power, and mirror orientation variables.

\vspace{-1mm}
\section{Performance Evaluation}\label{Sec:Results}
\vspace{-1mm}
\makeatletter
{\setlength{\@fptop}{0.75in}
\begin{algorithm}[t]
\small
\caption{Multi-Agent DRL for Joint RF/VLC Selection,
Power Allocation, and IRS Mirror Orientation}
\label{alg:maddpg}
\begin{algorithmic}[1]
\setlength{\baselineskip}{0.95\baselineskip}
\State \textbf{Initialization:}
\State Initialize actors $\mu_{SP}$, $\mu_{IRS}$ and critics
$Q^{\mu}_{SP}$, $Q^{\mu}_{IRS}$ with random parameters
$\theta^{\mu_{SP}}$, $\theta^{\mu_{IRS}}$, $w_{SP}$,
$w_{IRS}$
\State Initialize target networks
$\theta^{\mu'_i} \!\leftarrow\! \theta^{\mu_i}$,
$w'_i \!\leftarrow\! w_i$,
$\forall i \in \{SP, IRS\}$
\State Initialize replay buffer $\mathcal{D}$, mini-batch
size $\mathcal{N}_b$
\State \textbf{Training:}
\For{$\text{episode} = 1$ \textbf{to} $N_{ep}$}
    \State Observe initial state $s_0$
    \For{$t = 1$ \textbf{to} $T$}
        \State $AG_{IRS}$ selects mirror orientation action:
        \Statex \quad $a_{IRS,t} = \big[\, \varphi_{m,t},\,
        \vartheta_{m,t} \;\big|\; m \in \mathcal{M} \,\big]$
        \State Update VLC channel gains
        $\forall k \in \mathcal{K}$:
        \Statex \quad $h_{k,t}^{(v)} = \sum_{l=1}^{N_l}
        \big(b_{k,l,t}\, h_{k,l}^{\mathrm{LoS}} +
        \sum_{m=1}^{M} h_{k,m,l}^{\mathrm{IRS}}
        (\varphi_{m,t},\vartheta_{m,t})\big)$
        \State $AG_{SP}$ selects assignment and power action:
        \Statex \quad $a_{SP,t} = \big[\, \alpha_{k,t},\,
        P_{k,t}^{(v)},\, P_{k,t}^{(r)} \;\big|\;
        k \in \mathcal{K} \,\big]$
        \State Execute $a_t = \{a_{SP,t}, a_{IRS,t}\}$,
        compute $R_{k,t}$ using~\eqref{TDR} and reward $r_t$
        via~\eqref{eq:reward}, observe $s_{t+1}$
        \State Store $\{s_t, a_t, r_t, s_{t+1}\}$ in
        $\mathcal{D}$
        \If{$|\mathcal{D}| \geq \mathcal{N}_b$}
            \State Sample $\mathcal{N}_b$ transitions
            $\{s^j, a^j, r^j, s'^j\}$ from $\mathcal{D}$
            \For{each agent $i \in \{SP, IRS\}$}
                \State Compute target value:
                \Statex \qquad $y_i = r_i + \gamma
                Q_i^{\mu'}\big(s', \mu'_{SP}(o'_{SP}),
                \mu'_{IRS}(o'_{IRS}); w'_i\big)$
                \State Update critic by minimizing:
                \Statex \qquad $\mathcal{L}(w_i) =
                \frac{1}{\mathcal{N}_b}
                \sum_{j=1}^{\mathcal{N}_b}
                \big(Q_i^{\mu}(s^j, a_{SP}^j, a_{IRS}^j;
                w_i) - y_i\big)^2$
                \State Update actor using policy gradient:
                \Statex \qquad $\nabla_{\theta^{\mu_i}} J \approx
                \frac{1}{\mathcal{N}_b} \sum_{j=1}^{\mathcal{N}_b}
                \nabla_{\theta^{\mu_i}} \mu_i(o_i^j)\,
                \nabla_{a_i} Q_i(\cdot)
                \big|_{a_i = \mu_i(o_i^j)}$
                \Statex \qquad\quad where $Q_i(\cdot) =
                Q_i(s^j, a_{SP}^j, a_{IRS}^j)$
                \State Soft update target networks:
                \Statex \qquad $\theta^{\mu'_i} \!\leftarrow\!
                \tau \theta^{\mu_i} \!+\!
                (1\!-\!\tau)\theta^{\mu'_i}$, \;
                $w'_i \!\leftarrow\! \tau w_i \!+\!
                (1\!-\!\tau)w'_i$
            \EndFor
        \EndIf
        \State $s_t \leftarrow s_{t+1}$
    \EndFor
\EndFor
\State \textbf{Output:} Trained policies $\mu^*_{SP}$,
$\mu^*_{IRS}$
\end{algorithmic}
\end{algorithm}}
\vspace{-1mm}
\subsection{ Simulation configuration}   
We consider the system model described in Section~\ref{Sec:system} in a $5 \times 5 \times 3$~m indoor environment. The LED APs are mounted on the ceiling with a half-power semi-angle $\Phi_{1/2} = 60^\circ$, an optical transmit power of $2~\mathrm{W}$, and a bandwidth of $30~\mathrm{MHz}$, while the RF AP $N_f = 1$ is deployed on a wall at $1.25~\mathrm{m}$ height with a bandwidth of 20~MHz. An IRS array of $M = 49$ mirrors is mounted on one wall, with mirror area $dA_m = 2 \times 2$ $\mathrm{{cm}^2}$, spacing $q_m = 1~\mathrm{cm}$, and reflectivity $\rho_m = 0.95$. Each receiver has a photodetector responsivity of 0.5~$\mathrm{A/W}$, optical filter gain of 1.0, concentrator refractive index of 1.5, and noise power spectral density of $2\times10^{-14}$~A$^2$/$\mathrm{Hz}$. $K = 6$ mobile users are uniformly distributed on the floor with a minimum QoS rate of $R_{\min} = 1$~$\mathrm{Mbps}$. Each user moves at a speed $v \sim \mathcal{U}(0,2)$~m/s following the RWP mobility model with zero pause time and a distinct blockage probability $p_{b,k} \sim \mathcal{U}(0.1,0.6)$ reflecting heterogeneous indoor shadowing. Moreover, the simulation is implemented in Python 3.12 with PyTorch 2.0 and trained over 10000 episodes of 100 time steps each. The hyperparameters are: discount factor $\gamma = 0.99$, soft update rate $\tau = 0.01$, buffer capacity $|\mathcal{D}| = 10^5$, and mini-batch size $\mathcal{N}_b = 256$. Each actor comprises two fully connected hidden layers of 256 and 128 neurons, while each critic takes the joint state $s_t$ and both agents' actions as input through an identical architecture. ReLU activations are used in all hidden layers, and Tanh at the actor output. All networks are trained using Adam with learning rates $lr_\mu = 10^{-4}$ and $lr_Q = 3\times10^{-4}$, with exploration provided by Ornstein-Uhlenbeck noise of standard deviation $\sigma = 0.15$. The penalty coefficients are chosen with $\lambda_1 > \lambda_2$, prioritizing QoS satisfaction. 

\subsection{ Simulation Results}

\begin{figure}[t]
    \centering
    \includegraphics[width=0.45\textwidth]{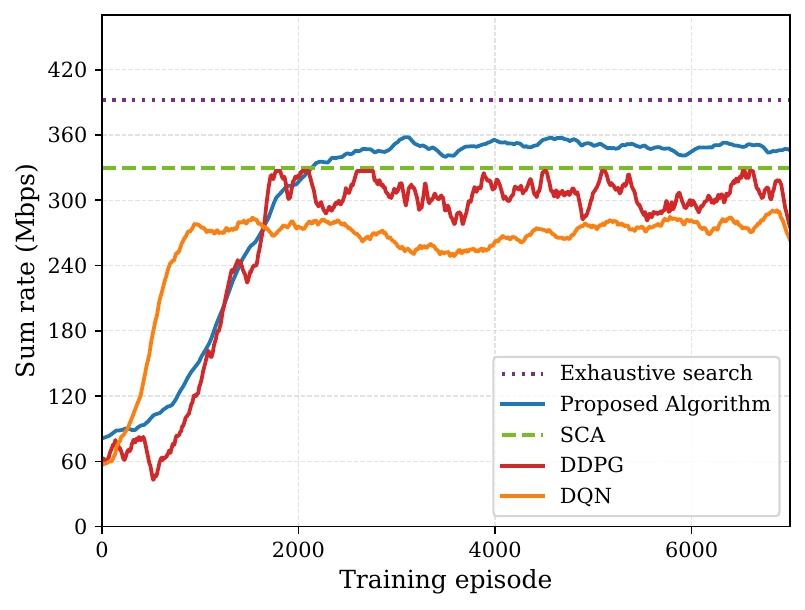}
    \caption{Sum rate versus episode during training.} 
    \vspace{-3mm}
    \label{fig:Fig2}
\end{figure}

The performance of our proposed algorithm is evaluated in terms of the sum rate over training episodes as shown in Fig.~\ref{fig:Fig2} for $K=6$ users. The proposed algorithm converges after approximately 2000 episodes to a stable sum rate of 350~Mbps, within 10\% of the exhaustive search upper bound of 392~Mbps. This confirms the near-optimal performance of our learned policy. The deterministic SCA method achieves 330~Mbps and must be re-executed at every scheduling slot. Moreover, our algorithm outperforms the DDPG and DQN by 13\% and 25\%, respectively. This is because DQN converges fastest due to its simpler discrete action space but saturates early at 280~Mbps, as it cannot capture the continuous action space. Additionally, the DDPG algorithm achieves a higher steady-state rate than DQN but oscillates by 25~Mbps around the SCA benchmark due to the instability of handling the hybrid discrete-continuous action space within a single agent, while the proposed multi-agent DRL algorithm remains stable with oscillations below 5~Mbps.

    \begin{figure}[t]
    \centering
    \includegraphics[width=0.45\textwidth]{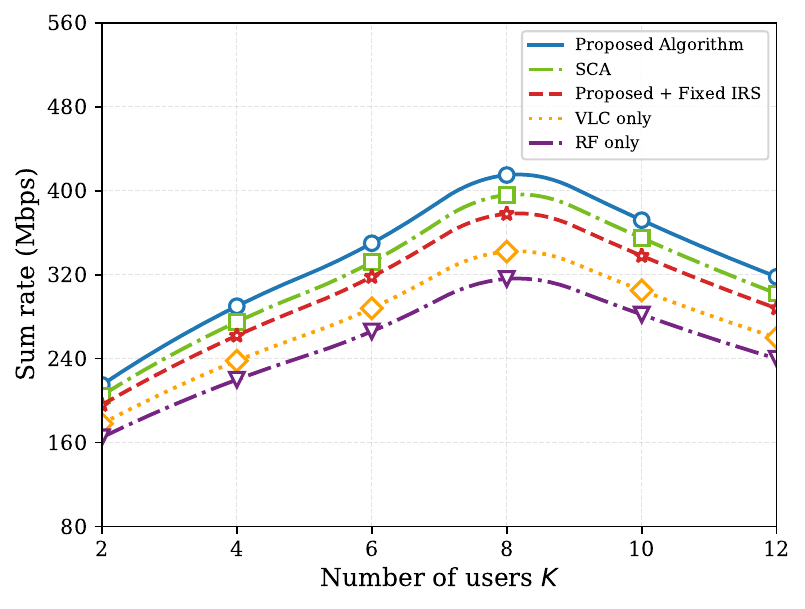}
    \caption{Sum rates versus different numbers of users.}%
    \vspace{-4mm}
    \label{fig:Fig3}
\end{figure}
\begin{figure}[t]
    \centering
    \includegraphics[width=0.45\textwidth]{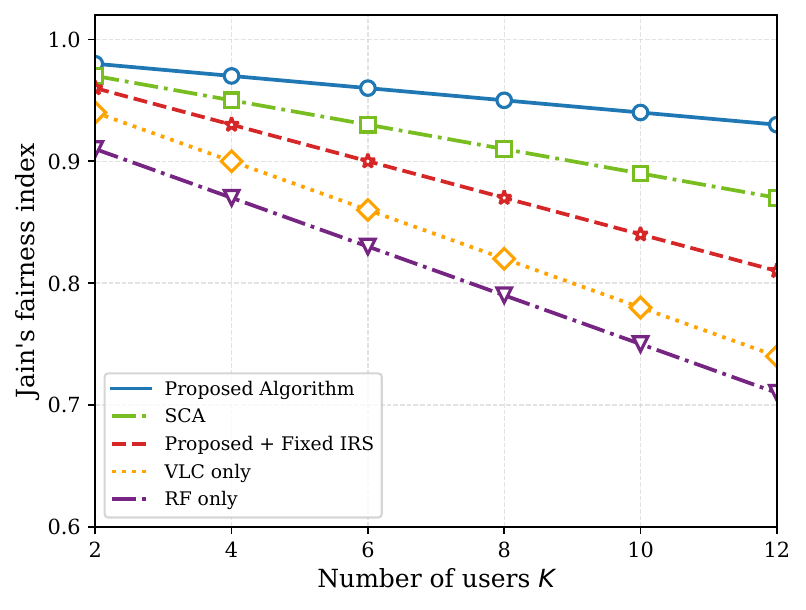}
    \caption{ Jain’s fairness index for different numbers of users.}
    \label{fig:Fig4}
    \vspace{-5mm}
\end{figure}

Fig.~\ref{fig:Fig3} shows the sum rate for different numbers of users. It can be seen that all approaches initially improve with $K$ until reaching a peak at $K=8$, beyond which the growing inter-user interference becomes the limiting factor. The proposed algorithm consistently outperforms all baselines, achieving a peak sum rate of 415~Mbps. This is because the multi-agents jointly adapt the technology selection, power allocation, and mirror orientation to the instantaneous channel conditions of each user. The deterministic SCA approach provides the second-best performance, as it optimizes all variables but converges to local optima rather than the global policy learned by the DRL agents. Moreover, disabling the IRS agent reduces the peak to 380~Mbps, which quantifies the contribution of dynamic mirror reconfiguration. 
Among the single-subnetwork baselines, VLC only outperforms RF only due to its larger bandwidth. However, both fall below the hybrid schemes. In other words, if a user experiences severe LoS blockage, VLC only cannot recover it without RF, while RF only cannot use the larger optical bandwidth available.

To evaluate the system fairness, Jain's fairness index \cite{11370966} is adopted as the performance metric. Fig.~\ref{fig:Fig4} presents the fairness of the
proposed algorithm against the same baselines for different numbers of users $K$. It can be seen that the fairness of all schemes decreases as $K$ increases, due to growing inter-user interference and the difficulty of meeting the QoS requirement in \eqref{C1} for a larger number of users. The proposed algorithm maintains the highest fairness index above 0.93 even at $K=12$. This is due to the use of multi-agents, where the one agent continuously steers the reflected beams toward users, while the other agent optimizes network selection and power allocation, thereby improving the rates of all users, including those in the worst conditions. Furthermore, SCA achieves fairness close to the proposed algorithm, as it also optimizes the mirror orientations alongside selection and power. In contrast, disabling the IRS agent reduces the fairness to 0.81 at $K=12$, since blocked users can no longer be recovered and their rates remain persistently low. Finally, VLC only and RF only result in the poorest fairness, as neither can exploit technology diversity to
serve users whose preferred subnetwork is temporarily unavailable.

\section{Conclusion} \label{Sec:Conc}
In this paper, a joint optimization problem was formulated for network selection, power allocation, and IRS mirror orientation in dynamic indoor hybrid RF/VLC networks to maximize the proportional fairness utility, which balances aggregate sum rate and inter-user fairness. Due to the non-convex mixed-integer nature of the optimization problem, it was reformulated as an MDP and solved using a multi-agent DRL algorithm operating under centralized training with decentralized execution, where one agent handles technology selection and power allocation while a second agent continuously reconfigures the IRS mirror angles in response to user mobility and LoS blockage. Simulation results demonstrated that the proposed algorithm consistently outperforms other baselines in sum rate and maintains a fairness index above 0.93 at high user densities. The results further confirm that the proposed multi-agent decomposition yields more stable convergence than traditional DRL algorithms.

\begingroup
\footnotesize
\setlength{\itemsep}{0pt}
\setlength{\parskip}{0pt}
\setlength{\parsep}{0pt}
\renewcommand{\baselinestretch}{0.9}
\bibliographystyle{IEEEtran}
\bibliography{References}
\endgroup
\end{document}